\documentclass[12pt,preprint]{aastex}
\usepackage{graphicx}

\shorttitle{Modeling the oxygen K absorption in the ISM}
\shortauthors{Garc\'{\i}a et al.}

\begin{document}
\title{Modeling the oxygen K absorption in the interstellar medium: an {\it XMM}-Newton view of Sco X-1}

\author{J.~Garc\'ia\altaffilmark{1,2},
        J. M.~Ram\'irez\altaffilmark{3},
        T. R.~Kallman\altaffilmark{2},
        M.~Witthoeft\altaffilmark{2},
        M. A.~Bautista\altaffilmark{1},
        C.~Mendoza\altaffilmark{4},
        P.~Palmeri\altaffilmark{5}, and
        P.~Quinet\altaffilmark{5,6}}

\altaffiltext{1}{Department of Physics, Western Michigan University, Kalamazoo, MI 49008, USA \email{javier.garcia@wmich.edu, manuel.bautista@wmich.edu}}
\altaffiltext{2}{NASA Goddard Space Flight Center, Greenbelt, MD 20771 \email{timothy.r.kallman@nasa.gov, michael.c.witthoeft@nasa.gov}}
\altaffiltext{3}{Astrophysikalisches Institut Potsdam, An der Sternwarte 16 D-14482 Potsdam, Germany \email{jramirez@aip.de}}
\altaffiltext{4}{Centro de F\'isica, IVIC, Caracas 1020, Venezuela \email{claudio@ivic.gob.ve}}
\altaffiltext{5}{Astrophysique et Spectroscopie, Universit\'e de Mons - UMONS, B-7000 Mons, Belgium \email{palmeri@umh.ac.be}}
\altaffiltext{6}{IPNAS, Sart Tilman B15, Universit\'e de Li\`ege, B-4000 Li\`ege, Belgium \email{quinet@umh.ac.be}}

%
\begin{abstract}
We investigate the X-ray absorption structure of oxygen in the
interstellar medium by analyzing {\it XMM}-Newton observations
of the low-mass X-ray binary Sco X-1. We use simple models
based on the O~{\sc i} atomic photoabsorption cross section
from different sources to fit the data and evaluate the impact
of the atomic data on the interpretation of the
observations. We show that relatively small differences in the
atomic calculations can yield spurious results, and
that the most complete and accurate set of atomic cross
sections successfully reproduce the observed data in the
$21.0{-}24.5$~{\AA} wavelength region of the spectrum. Our fits
indicate that the absorption is mainly due to neutral gas with
an ionization parameter of $\xi=10^{-4}$~erg~cm~s$^{-1}$ and
an oxygen column density of $N_{\mathrm{O}}\approx (8{-}10)\times
10^{17}$~cm$^{-2}$. Our models are able to reproduce both the K
edge and the K$\alpha$ absorption line from O~{\sc i} which
are the two main features in this region. We find no conclusive
evidence for absorption by other than atomic oxygen.
\end{abstract}
%
%
\section{Introduction}
X-ray spectroscopy provides a powerful tool for understanding the physical and
chemical properties of the diffuse interstellar medium (ISM). The X-ray band covers
the emission and absorption spectra produced by inner-shell transitions of
the most abundant ions from carbon to iron. The interaction of X-ray photons
from bright point sources (e.g. galactic X-ray binaries) with the ISM gives rise
to absorption lines and edges in the spectrum.
The energy position and the shape of these features depend on whether the
absorption is due to free atoms or molecules, and on whether these atoms or
molecules are in the gas or in the solid phase.

Neutral oxygen is a major constituent of the ISM which makes
it one of the most important elements in astronomical observations. Precise
knowledge of the neutral oxygen atomic parameters is needed for the correct
modeling of the observed spectra. Calculations of the photoabsorption
cross section of the ground state of O~{\sc i} were carried out by \cite{mcl98}
(hereafter {\sc mck98}) using the {\it R}-matrix method, giving a detailed
comparison with the experimental results of \cite{sto97}. Although they claimed
overall agreement, there are significant discrepancies in the positions
of the inner-shell excited resonances and in the near-threshold resonance profiles.
This problem was overcome in the $LS$-coupling calculation of \cite{gor00}
(hereafter {\sc gmc00}) by taking into account core relaxation effects and the
smearing of the K edge due to Auger damping. A more complete {\it R}-matrix
calculation was carried out in intermediate coupling by \cite{gar05} (hereafter
{\sc gar05}) for all the ions in the oxygen isonuclear sequence. There is
now very good agreement between the {\sc gar05} calculations and both the
experimental cross section \citep{sto97} and the {\sc gmc00} results.

The oxygen inner-shell features in the X-ray spectrum of galactic sources have been
used to provide abundance determinations in the ISM as well as estimates of the
oxygen ionization fractions \citep{sch02,tak02,jue04,tur04,ued05,jue06}. However,
studies in the IR and UV have shown that oxygen can also be found in solid
particles \citep{dra03,whi03}. It has been argued that oscillatory modulations
near the K edge, usually referred to as the X-ray absorption fine structure
(XAFS), could be detected. These are condensed matter modulations of the atomic cross section
due to the presence of solid particles \citep{lee05,lee09}. Studies of the soft
X-rays from galactic sources have reported possible detections of molecules that
could be linked to XAFS signatures in the edges of several elements such as
Ne, Si, and Mg \citep{pae01,lee02,ued05}; in the L edge of Fe \citep{lee01,kas09};
and in particular, in the oxygen K edge \citep{dev03,cos05,dev09,pin10}.
Although these signatures could be important in the K edges of molecules involving higher
$Z$ elements, namely Mg, Si, and Fe, oxygen is $\sim 10{-}20$ times more abundant than
any of these, thus potentially providing enough signal-to-noise to detect the XAFS signatures.
Theoretical models of XAFS in the astrophysical context have been developed by \cite{woo95,woo97};
and \cite{for98}. See also \cite{lee11} and references therein for details
on the theory of XAFS.

{\it XMM}-Newton observations of the X-ray source Scorpius~X-1 (Sco~X-1)
reveal strong absorption in the wavelength region corresponding to neutral
oxygen. Located at a $\sim 2.8$~kpc distance \citep{bra99} and with a
flux of $F\sim 3.4\times 10^8$~erg~cm$^{-2}$~s$^{-1}$ (in the $2{-}10$~keV energy
band), it is the brightest X-ray source in the sky other than the Sun and
the diffuse X-ray background radiation. Its high X-ray flux provides very
good statistics in relatively short exposure times, giving the opportunity
to study signatures of oxygen absorption in the ISM with great detail. For
several galactic sources, including Sco~X-1, \cite{devr03} analyzed
high-resolution X-ray spectra taken with the reflection grating spectrometer
(RGS) in the {\it XMM}-Newton satellite. By comparing low and high
extinction sources, they were able to separate the ISM and the instrumental
components of the O~{\sc i} K edge; moreover, \cite{dev09} searched for XAFS
signatures in the spectrum of Sco~X-1. The XAFS signature is derived from
the differences between the observed flux and that predicted theoretically.
However, the model used by these authors is based on the atomic oxygen
absorption cross section calculated by {\sc mck98}.

In this Letter we show the importance of the accuracy of the atomic data
used in the modeling of the detailed features of the oxygen absorption in the ISM. We
demonstrate that small variations in the K-edge structure derived from different
atomic calculations yield spurious results when applied to astronomical observations.
In Section~\ref{secobs}, we describe the observational data used in our study
while the theoretical models are delineated in Section~\ref{secmod}. Results
derived from model fits of the observed data are presented in Section~\ref{secres}, and
finally, the main conclusions are summarized in Section~\ref{seccon}.
%
%
\section{Observation and data reduction}\label{secobs}
For the purpose of this work, we adopt the {\it XMM}-Newton spectrum of Sco~X-1
taken in orbit 0592 (Obs ID = 0152890101) with the RGS instrument \citep{den01}.
The general observational strategy is described in detail in \cite{devr03} and
\cite{dev09} and will not be given here. Taking into account the high X-ray flux of Sco~X-1,
standard spectroscopy mode would lead to a very high pile-up level. This was avoided
by choosing a faster readout mode which is able to read each of the nine RGS
CCDs in separate exposures. Because of a failure in the RGS2, we make use of only
the RGS1 data in what follows. During this observation there were 24 separate
contiguous exposures, but only the fifth and the eighteenth contained calibrated
and reliable data in the $21{-}25$~{\AA} spectral range; thus, we only use these exposures
in our spectral analysis. In the notation of the SAS {\tt rgsproc} task,
these two exposures are labeled S005 and S018. We follow the standard procedure
for the reduction and extraction of the RGS spectrum using {\sc SAS}~version 10
with the latest calibration files (CCFs). We finally present the spectrum
with the default binning of 0.05 \AA\ employing full spectrometer resolution.
%
%
\section{Spectroscopic models}\label{secmod}
\subsection{Atomic data}
\begin{figure*}
\epsscale{1.0}\plotone{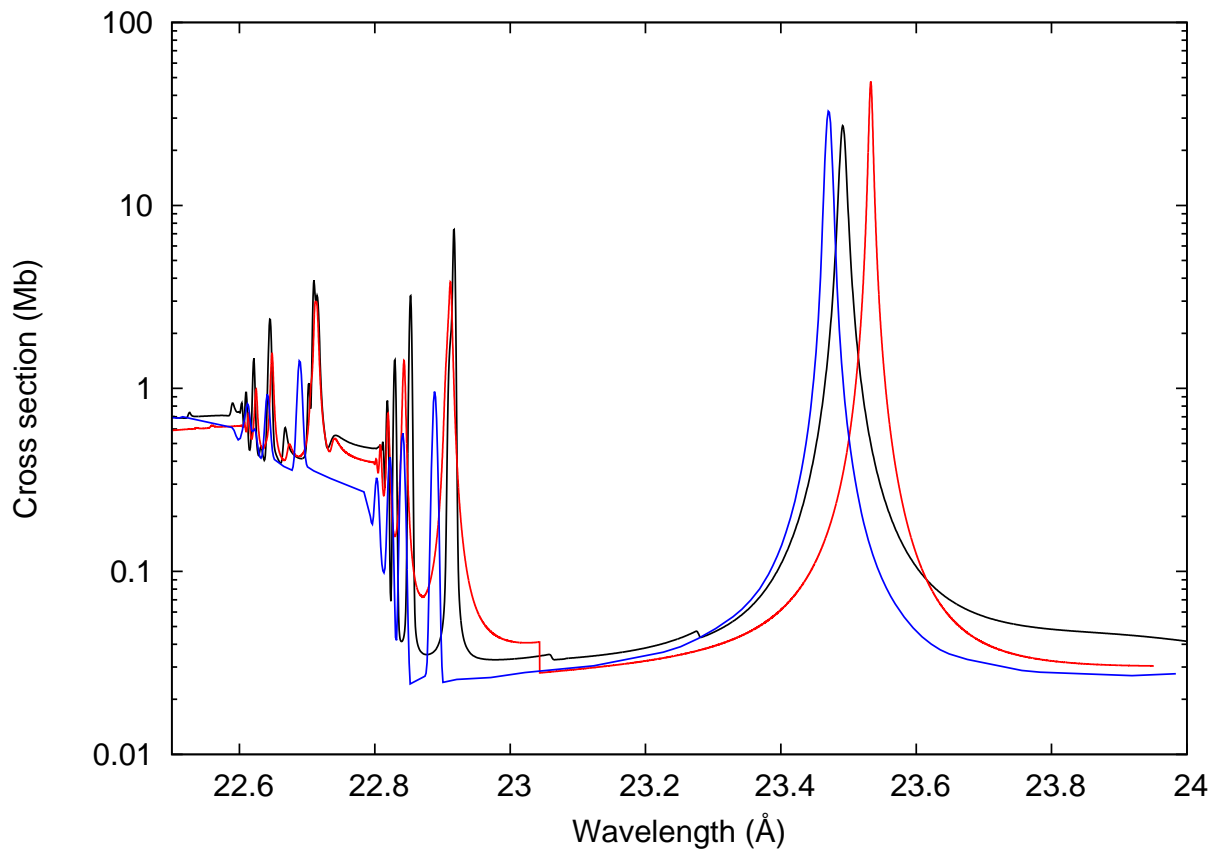}
\caption{Comparison of the theoretical photoabsorption cross sections of neutral oxygen
in the $22.5{-}24.0$~{\AA} wavelength region: \cite{mcl98} (blue curve); \cite{gor00} (red curve);
and \cite{gar05} (black curve). This spectral region covers both the absorption K edge and the K$\alpha$
absorption line (${\rm 1s}{-}{\rm 2p}$) from O~{\sc i}. All the curves have been convoluted with
a $0.182$~eV FWHM Gaussian profile.}
\label{f1}
\end{figure*}

Figure~\ref{f1} shows the photoabsorption cross section of the ground state
of neutral oxygen in the $22.5{-}24.0$~\AA\ wavelength region from the calculations by
{\sc mck98}, {\sc gmc00}, and {\sc gar05}. All these curves have been convolved
with a Gaussian profile of $0.182$~eV full width at half maximum (FWHM) in
order to match the resolution of those presented in {\sc gmc00}. This spectral region covers
both the K edge around 22.5~{\AA} and the K$\alpha$ ${\rm 1s}{-}{\rm 2p}$ absorption line near 23.5~\AA.
Besides the differences in the K$\alpha$ position, the {\sc gar05} and {\sc gmc00} calculations
agree very well. Nevertheless, there are significant discrepancies with {\sc mck98},
particularly in the shape of the K edge and in the energy separation between the K edge and
the K$\alpha$ resonance. Since the K$\alpha$ absorption feature is prominent and well-resolved
and its energy is experimentally fixed, it is used as a reference in most
spectral fits. Thus, uncertainties in the energy and shape of the cross section
around the inner-shell edges are translated in the form of spurious residuals in the fits.

These discrepancies have been formally addressed by {\sc gmc00} as the result to two
conspicuous effects: orbital relaxation due to the K shell vacancy which affects the
atomic structure, and consequently, the resonance and edge positions; and resonance decay
through an infinite number of transitions wherein the resonance width is usually dominated
by the spectator-Auger channel. It is evident from the comparison in Figure~\ref{f1} that
the neglect of these two effects underestimates the absorption cross section
near the K edge by $20{-}50$\%.
\subsection{Model A}\label{secmoda}
In order to resolve the features in the Sco X-1 observation, we have employed the
photoionization code {\sc xstar}. Several calculations were performed covering
a range of parameters, the most important being the hydrogen column density,
$N_{\mathrm{H}}$, and the ionization parameter, $\xi = L/nR^2$, where $L$ is the
luminosity of the source, $R$ its distance, and $n$ is the density of the gas \citep{tar69}.
We have constructed a grid of {\sc xstar} models spanning hydrogen column densities
of $N_{\mathrm{H}}=10^{19}{-}10^{22}$ cm$^{-2}$ and ionization parameters of $\xi=10^{-4}{-}10$
erg cm$^{-2}$ s$^{-1}$ with the gas density fixed at $n=10^3$~cm$^{-3}$. The
spectral region of interest, $21.0{-}24.5$~{\AA}, is relatively small and, in practice,
is dominated by oxygen species. Thus, the {\sc xstar} models include hydrogen, helium, and oxygen
ions in the ionization balance calculation, assuming an oxygen abundance relative to hydrogen of
$A_{\mathrm{O}}=6.8\times 10^{-4}$ \citep{gre98}. We will refer to this model as Model~A.
The {\sc xstar} models incorporate the {\sc gar05} cross sections for all the oxygen charge states.
\subsection{Models B, C and D}\label{secmodbcd}
If a simple description of X-ray absorption by a cold medium assumes that all
the spectral features are solely due to neutral oxygen photoabsorption,
the observed flux can then be approximated as
\begin{equation}\label{eq1}
F(E) = F_0 \exp\left[-N_{\mathrm{O_I}}\sigma_{\mathrm{O_I}}(E)\right]
\end{equation}
where $F_0$ is a normalization factor, $N_{\mathrm{O_I}}$ is the oxygen column
density, and $\sigma_{\mathrm{O_I}}(E)$ the photoabsorption cross section
of O~{\sc i}. This is a convenient way to evaluate
the relevance of the atomic data used to fit the observation. Using this assumption,
we have produced three additional models using Equation~(\ref{eq1}), each
with a different cross section: Model~B includes that computed by {\sc mck98};
Model~C by {\sc gar05}; and Model~D by {\sc gmc00}.
Note that the {\sc xstar} model (Model~A) and Model~C are equivalent
in the sense that they use the same atomic data although the former
also includes the background due to H and He. These models are
summarized in Table~\ref{tamodels}.
%
%
\section{Results}\label{secres}
\begin{figure*}
\epsscale{1.0}\plotone{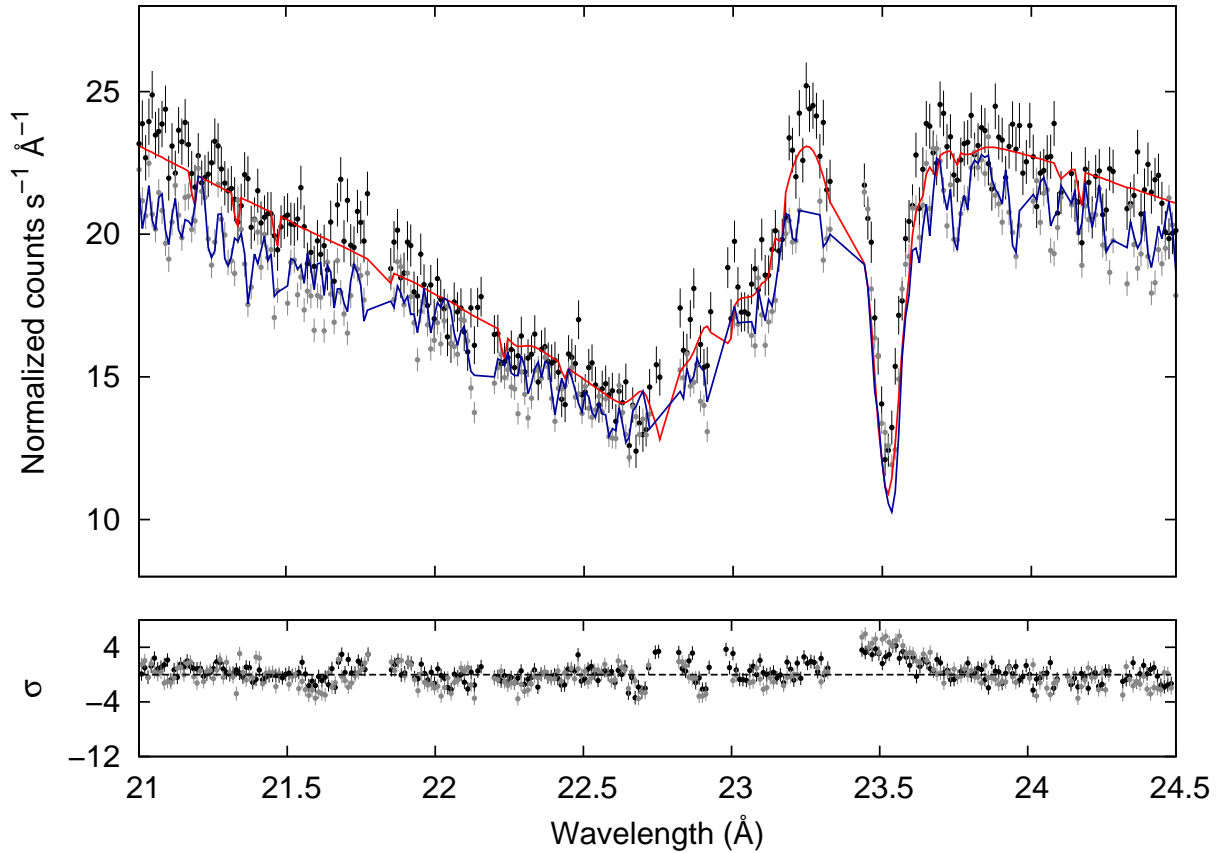}
\caption{The Sco X-1 spectrum produced by RGS1 aboard {\it XMM}-Newton covering the $21.0{-}24.5$~{\AA}
wavelength region. In the upper panel, the black and gray data points correspond to exposures S005 and S018,
respectively, while the best fits to each exposure using the {\sc xstar} photoionization model (Model~A)
are respectively depicted in red and blue. The lower panel shows the residuals in $\sigma$ units in the
same range with respect to the model. Black and gray points correspond to exposures S005 and S018,
respectively.}
\label{f2}
\end{figure*}

The models described in Section~\ref{secmod} are used to fit the
absorption features observed in the X-ray spectrum of Sco~X-1. In this respect and to
determine the corresponding statistics, we use the X-ray spectral
package {\sc xspec}~v12.3.0, and all the fits are carried out in the
$21.0{-}24.5$~{\AA} spectral region. Figure~\ref{f2} shows the fit using Model~A,
our main model, since it is the result of a self-consistent photoionization calculation
and incorporates the most recent atomic cross section for O~{\sc i} by {\sc gar05}.
In the upper panel, the black and gray data points belong to exposures
S005 and S018, respectively. The best fit using Model~A is shown with solid lines,
red and blue respectively corresponding to the fit applied to each
exposure. The spectrum displays a very strong atomic K edge which
covers the $22.8{-}23.3$~{\AA} wavelength range. The K$\alpha$
absorption resonance at $\sim 23.5$~{\AA} is also one of the strongest features
while the K$\beta$ is much weaker but is still detectable
at $\sim 22.9$~\AA. In the lower panel, we show the residuals with respect to
the model in $\sigma$ units. The black and gray points correspond to exposures
S005 and S018, respectively. Model~A fits the K edge and the K$\alpha$
absorption line successfully in the two exposures of Sco~X-1. Statistics for the
combined fit (i.e., for the two exposures) show a reduced chi-square of
$\chi^2/{\mathrm{dof}}=2.75620$ (where dof is the number of degrees of freedom).
The best-fit hydrogen column density is
$N_{\mathrm{H}}=(1.33\pm0.02)\times 10^{21}$~cm$^{-2}$ which corresponds to an oxygen column
density of $N_{\mathrm{O}}=(9.04\pm0.12)\times 10^{17}$~cm$^{-2}$.

\begin{figure*}
\epsscale{1.0}\plotone{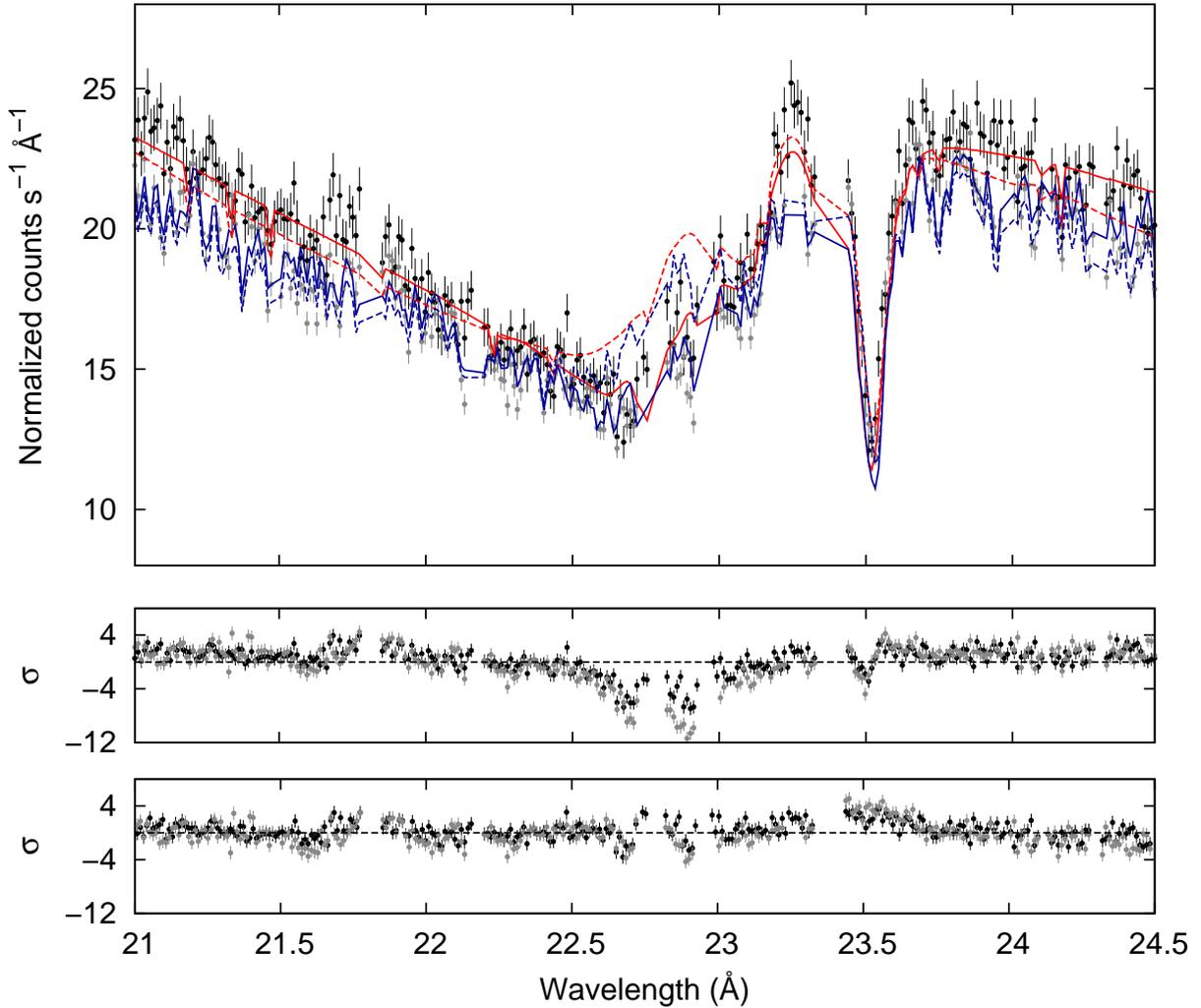}
\caption{Spectral fit comparison using two different models. Upper panel: Sco X-1
spectrum covering the $21.0{-}24.5$~{\AA} wavelength region. The black/gray data points
are the observed data while the red/blue curves are the models corresponding to exposures
S005/S018, respectively. The best fits using Models~B and C are shown with the dashed and
solid lines, respectively. Middle panel: residuals in $\sigma$ units with respect to
Model~B. Lower panel: residuals in $\sigma$ units with respect to Model~C.}
\label{f3}
\end{figure*}

In order to get a better grasp of the atomic data effects on the
description of the observed spectra, we now fit the measurements
with the simple models based on the raw photoabsorption cross section of neutral oxygen
(see Equation~\ref{eq1}). In Figure~\ref{f3} we depict the fits using Models~B and C; in
the upper panel, the black/gray data points are the measurements while the red/blue curves
are the models corresponding to exposures S005/S018, respectively. The best fits using
Models~B and C are respectively shown with dashed and solid lines. The middle panel
gives the residuals in $\sigma$ units with respect to Model~B and the lower panel
those with respect to Model~C. It is important to note that, if these two models
were equivalent, the dashed and solid lines with the same color would be
close to each other. However, there is a clear discrepancy between the dashed
lines (Model~B) and the solid lines (Model~C) in the region near the oxygen K edge
($\sim 22.5{-}23$~\AA). These differences may be appreciated in the residuals
of the Model~B fit shown in the middle panel. Furthermore, the residuals also indicate
that Model~B cannot completely fit the intensity of the K$\alpha$ absorption
line at $\sim 23.5$~\AA; it predicts an oxygen column density of
$N_{\mathrm{O}}=(9.25\pm 0.07)\times 10^{17}$~cm$^{-2}$; and the fit statistics give
a reduced chi-square of $\chi^2/{\mathrm{dof}}=6.02484$ which is poorer than Model~A.
As expected, the fit using the raw cross section by {\sc gar05} (Model~C) is equivalent to
that using Model~A giving a reduced chi-square of $\chi^2/{\mathrm{dof}}=2.44650$.
Model~C predicts an oxygen column density of $N_{\mathrm{O}}=(7.94\pm 0.22)\times 10^{17}$~cm$^{-2}$
somewhat smaller than Models~A and B. The differences between Models~A and C
may be due to the numerical interpolation used in the storage and retrieval of
the cross sections by the {\sc xstar} package.

Model~B is the same as that used by \cite{dev09} to fit the oxygen absorption in the spectrum of
Sco~X-1. These authors claimed the detection of XAFS signatures in the
spectra based on the relative changes in the observed flux with respect to
the smooth flux predicted by the model for energies above the K edge. They
also argued that the apparent shift of the observed edge with respect to the
atomic model could be due to the fact that some fraction of the oxygen in
the ISM is bound in solids. However, the analysis presented here shows that
the large residuals found in the Model~B fit are artifacts brought about by
the {\sc mck98} cross section.

\begin{figure*}
\epsscale{1.0}\plotone{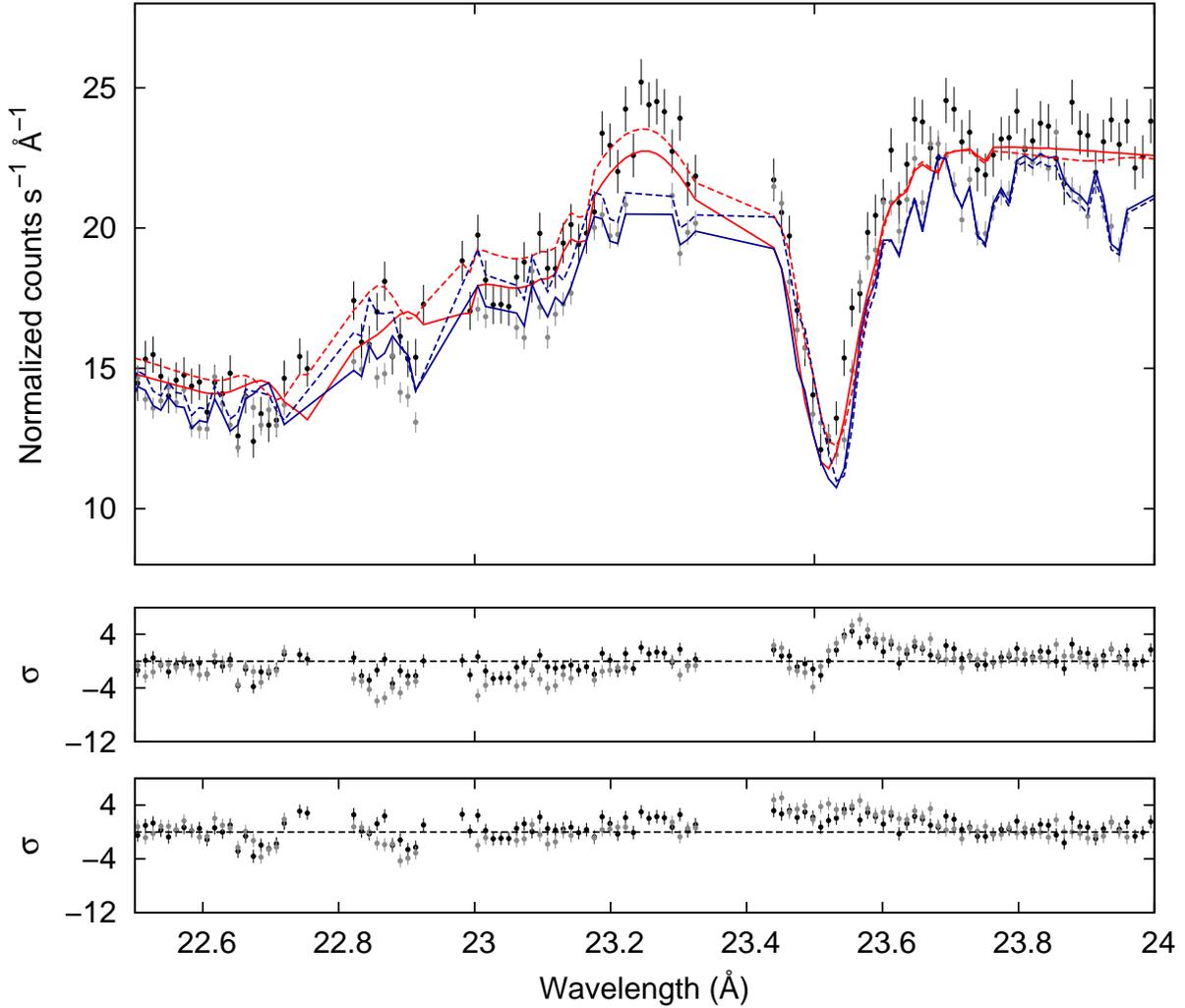}
\caption{Spectral fit comparison using two different models. Upper panel: Sco X-1 spectrum
in the $22.5{-}24.0$~{\AA} wavelength region. The black/gray data points are the observed data
while the red/blue curves are the models to exposures S005/S018, respectively. The best fits
using Models~C and D are shown with solid and dashed lines, respectively. Middle panel:
residuals in $\sigma$ units with respect to Model~D. Lower panel: residuals in $\sigma$ units
with respect to Model~C.}
\label{f4}
\end{figure*}

Finally, in Figure~\ref{f4} we show a similar comparison using the {\sc gar05}
and the {\sc gmc00} cross sections (Models~C and D). As before, the upper
panel shows the Sco X-1 spectrum and the best-fit models. The black/gray data
points are the observations while the red/blue curves are the models corresponding
to exposures S005/S018, respectively. Note that Figure~\ref{f4} covers the
$22.5{-}24.0$~{\AA} wavelength range to enhance the K-edge region. The best fits
using Models~C and D are respectively shown with solid and dashed lines. The
middle and lower panels give the residuals in $\sigma$ units for Models~D and C. The
two models are equivalent giving similar fits ($\chi^2/{\mathrm{dof}}=2.76445$
for Model~D). This is consistent with the accord between the {\sc gmc00} and {\sc gar05}
cross sections. The oxygen column density derived from this model,
$N_{\mathrm{O}}=(10.49\pm 0.06)\times 10^{17}$~cm$^{-2}$, is larger than those
estimated by all the previous models, but consistent within the uncertainties in the
parameters predicted by Models~A and C. The discrepancies in the predicted oxygen columns
in Models~C and D can be explained by those in their respective cross sections.
In this small spectral region, the oxygen column depends almost entirely on the
depth of the K edge. By comparing the values of the cross section
at $22.8$ and $23.0$~{\AA} (i.e. before and after the K edge) in Figure~\ref{f1},
it may be noticed that the K edge in {\sc gmc00} is weaker than in
{\sc gar05}. If a feature is weaker in a model, it will require a larger column
density to reproduce the observation and, therefore, explains the different
column densities obtained. All the proposed models and the parameters derived from their
corresponding fits are summarized in Table~\ref{tamodels}.

In all the fits presented here, we see residuals of around $\pm 2 \sigma$ distributed
homogeneously along the entire spectral range considered in our analysis, consistent
with the reduced chi-square close to 2 obtained in our best fit. This suggests that
most of the fit errors are systematic. However, we notice significant residuals
at wavelengths shorter than the K$\alpha$ line. Due to a gap in the $23.33{-}23.44$~{\AA}
wavelength range, the detection of spectral features is not possible; nonetheless,
both the atomic K$\alpha$ resonance due to ionized oxygen and the $1s\rightarrow \pi^*$
resonance due to molecular oxygen occur at $23.35$~{\AA}. The former alternative is
consistent with the analysis by \cite{jue04} where they detected O~{\sc ii} absorption
in the spectra of several X-ray binaries. On the other hand, our fit using Model~A
predicts that absorption occurs in a mostly neutral gas with an ionization parameter
of $\xi=10^{-4}$~erg~cm~s$^{-1}$ which would also suggest molecular absorption.
Therefore, given the lack of data in this particular spectral range, we cannot rule
out the presence of molecules in the observation in hand.

%
%
\section{Conclusions}\label{seccon}
In the present report we have shown the relevance of accurate atomic data in the modeling
of the X-ray spectra from cosmic sources. In particular, we have studied the Sco X-1
spectrum produced by the RGS1 instrument on board of the {\it XMM} Newton
satellite covering the $21.0{-}24.5$~{\AA} wavelength region. Absorption occurs when
the X-rays interact with the cold gas of the ISM, the main spectral
features in this region being the absorption K edge and K$\alpha$ line from
neutral oxygen. We found a good fit using a self-consistent photoionization model
which includes the most recent atomic data for the oxygen isonuclear sequence by
{\sc gar05}. Our fits indicate that the absorbing gas has an ionization parameter
of $\xi=10^{-4}$~erg~cm~s$^{-1}$ and a hydrogen column density of
$N_{\mathrm{O}}\approx (8{-}10)\times 10^{17}$~cm$^{-2}$.

Simple models based on the raw atomic photoabsorption cross sections of O~{\sc i}
from three different calculations were used to evaluate data sensitivity. We
show that models based on the {\sc mck98} atomic cross sections are unable to reproduce
the K-edge structure in detail, while those based on {\sc gmc00} and {\sc gar05}
yield more accurate fits of the main spectral features. The fits using the most up to
date models do not show evidence for absorption by anything other than atomic oxygen.

The analysis presented here indicates that the atomic data uncertainties in combination
with the limited resolution of the grating spectrum make detection of molecular or solid
material challenging. Although oxygen is expected to be found in molecular form or locked
into solids in the ISM, the use of accurate atomic calculations to correctly account for
the atomic oxygen contribution is crucial when searching for XAFS or similar features
in the X-ray spectra of astronomical sources.
%
\acknowledgments
We thank Tom Gorczyca for providing the {\sc mck98} and {\sc gmc00} calculations.
This work was supported by a grant from the NASA astrophysics theory program
05-ATP05-18. This research has made extensive use of the NASA Astrophysics Data System.
%
%

%
%
\begin{deluxetable}{clcccl}
\tabletypesize{\scriptsize}
\tablecaption{List of Models
\label{tamodels}}
\tablewidth{0pt}
\tablehead{
\colhead{Model} & \colhead{Description} & \colhead{Atomic Data} & \colhead{$N_{\mathrm{O}}$} & \colhead{$\chi^2$/dof} & \colhead{Notes} \\
                &                       &                       & \colhead{($10^{17}$ cm$^{-2}$)}  &  &  \\
}
\startdata
 A  &  Full {\sc xstar} model                                   &
{\sc aka01}\tablenotemark{a}+{\sc gar05}\tablenotemark{b} & $9.04\pm 0.12$ & 2.75620 & No significant residuals \\
 B  &  Atomic cross section\tablenotemark{e} & {\sc mck98}\tablenotemark{c} & $9.25\pm 0.07$ & 6.02484 & Large residuals near K edge \\
 C  &  Atomic cross section\tablenotemark{e} & {\sc gar05}\tablenotemark{b} & $7.94\pm 0.22$ & 2.44650 & No significant residuals \\
 D  &  Atomic cross section\tablenotemark{e} & {\sc gmc00}\tablenotemark{d} & $10.49\pm 0.06$ & 2.76445 & No significant residuals \\
\enddata
\tablenotetext{a}{\cite{bau01}}
\tablenotetext{b}{\cite{gar05}}
\tablenotetext{c}{\cite{mcl98}}
\tablenotetext{d}{\cite{gor00}}
\tablenotetext{e}{See Equation~(\ref{eq1})}
\end{deluxetable}
%
%
%

%
%
%
%
%
\end{document}